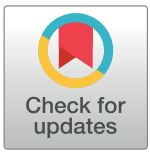

RESEARCH ARTICLE

# Use of a humanoid robot for auditory psychophysical testing

Luke Meyer[1,2]*, Laura Rachman[1,2], Gloria Araiza-Illan[1,2], Etienne Gaudrain[3], Deniz Başkent[1,2]

**1** Department of Otorhinolaryngology, Head and Neck Surgery, University Medical Center Groningen, University of Groningen, Groningen, The Netherlands, **2** W.J. Kolff Institute for Biomedical Engineering and Materials Science, University Medical Center Groningen, University of Groningen, Groningen, The Netherlands, **3** Lyon Neuroscience Research Center, CNRS UMR 5292, INSERM UMRS 1028, Université Claude Bernard Lyon 1, Université de Lyon, Lyon, France

* l.meyer@rug.nl

## Abstract

Tasks in psychophysical tests can at times be repetitive and cause individuals to lose engagement during the test. To facilitate engagement, we propose the use of a humanoid NAO robot, named Sam, as an alternative interface for conducting psychophysical tests. Specifically, we aim to evaluate the performance of Sam as an auditory testing interface, given its potential limitations and technical differences, in comparison to the current laptop interface. We examine the results and durations of two voice perception tests, voice cue sensitivity and voice gender categorisation, obtained from both the conventionally used laptop interface and Sam. Both tests investigate the perception and use of two speaker-specific voice cues, fundamental frequency (F0) and vocal tract length (VTL), important for characterising voice gender. Responses are logged on the laptop using a connected mouse, and on Sam using the tactile sensors. Comparison of test results from both interfaces shows functional similarity between the interfaces and replicates findings from previous studies with similar tests. Comparison of test durations shows longer testing times with Sam, primarily due to longer processing times in comparison to the laptop, as well as other design limitations due to the implementation of the test on the robot. Despite the inherent constraints of the NAO robot, such as in sound quality, relatively long processing and testing times, and different methods of response logging, the NAO interface appears to facilitate collecting similar data to the current laptop interface, confirming its potential as an alternative psychophysical test interface for auditory perception tests.

## Introduction

In social robotics, the main mode of communication with humans is speech [1]. In this study, we take advantage of the speech and communication tools on a low-cost humanoid robot, NAO, to conduct psychophysical tests on voice perception. From this implementation, three points of discussion can be derived: can a humanoid robot be used as an alternative interface to a computer implementation for psychophysical testing?, what is the perception people have

**Data Availability Statement:** All processed data files are available from the DataverseNL repository, DOI: https://doi.org/10.34894/IAGXVF.

**Funding:** D. Baskent: VICI Grant from the Netherlands Organization for Scientific Research (NWO) and the Netherlands Organization for Health Research and Development (ZonMw) (Grant No. 918-17-603) – https://www.nwo.nl/en; The funders had no role in study design, data collection and analysis, decision to publish, or preparation of the manuscript. L. Meyer, L. Rachman, G. Araiza-Illan, D. Baskent: W.J. Kolff Institute, University Medical

Centre Groningen (no grant number) – https://umcgresearch.org/w/wjkolff; The funders had no role in study design, data collection and analysis, decision to publish, or preparation of the manuscript. L. Meyer, L. Rachman, G. Araiza-Illan, E. Gaudrain, D. Baskent: Heinsius-Houbolt Foundation (No grant number); The funders had no role in study design, data collection and analysis, decision to publish, or preparation of the manuscript.

**Competing interests:** The authors have declared that no competing interests exist.

towards having a social robotic agent as an alternative interface and conduct clinical tests?, and how does the presentation of different voices, either natural or synthesised, potentially influence communication between a human and robot? The factors that influence these different components are vast; e.g., personality, education [2], background [3], gender [4] etc., and while these factors may have multiple interrelated connections between them, the focus of the current manuscript is on the first component; can a humanoid NAO robot be effectively used as a psychophysical testing interface?

Speaker-specific voice cues, such as fundamental frequency (F0), related to the perceived pitch of a voice and determined by glottal pulse rate, and vocal-tract length (VTL), related to the speech spectral profile and the size of the talker [5, 6], are two key characteristics in differentiating voices and identifying a speaker's voice gender [7–9]. Individuals with normal hearing are sensitive to voice cues, and can perceive very small differences, around 1–2 semitones [8, 10–12]. In multi-talker situations, the categorisation of voice based on gender may assist one in identifying and focussing on a voice, especially during simultaneous talking [7]. Furthermore, the categorisation of voice gender is not only important from a medical perspective, such as in the advancing and developing of devices utilised by hard-of-hearing individuals, but also in the context of social robot implementations where the perceived gender of either a natural or synthesised voice presented by a robot can influence its effectiveness, depending on its application. This is explored in an extensive meta-analysis by [13]. When categorising the perceived gender of a voice, normal-hearing listeners use both voice cues effectively [7, 8, 11, 14]. In contrast, hard-of-hearing users of auditory prosthetic devices; e.g, cochlear implants, have difficulty differentiating voice cues [8] and seem to rely heavily on F0 differences for voice gender categorisation while being unable to make use of VTL differences [7]. This example indicates much is still to be uncovered regarding the perception of voice and speech, especially with hearing devices, to fully understand the abilities and limitations of voice perception, and to accordingly improve performance of augmentative devices, such as hearing aids and cochlear implants.

In investigating voice and speech perception, the psychophysical tests are often long and repetitive to ensure data reliability [15–17]. Establishing and maintaining engagement and focus during such studies can be a challenge for all participants, but also especially for individuals with relatively short attention spans such as children [18, 19], or listeners with limited hearing abilities, who also often happen to be older individuals [20]. These populations are often understudied for voice perception, perhaps partially due to such challenges. Voice perception in children and younger and older adults with hearing loss has mostly recently started to be investigated [12, 21, 22].

The standard setup for auditory psychophysical tests involves a computer interface used for both the presentation of stimuli and collection of responses, which can be recorded in audio or through a simple interaction involving mouse clicks, keyboard entry or similar. To help with potential attention and engagement issues and to maintain good quality of test results, computer interfaces are at times further modified to include cartoons, cartoon characters, or animations [12, 23, 24]. In this study, as a new alternative psychophysical test interface to the conventional computer interface, we propose a humanoid NAO robot, which could potentially result in the collection of sufficiently reliable measurements of perceptual performance while doing so in an engaging manner. As suggested by [1], speech with a robot can be advantageous in motivating and engaging users. Further supporting this, [25] have shown that between a humanoid robot and a computer, the robot was better at retaining the attention of children during learning tasks. The literature also shows that the physical embodiment of an agent is preferred over its virtual counterpart, and contributes favourably to its social presence [26–29], potentially motivating interactors to exert more effort on a given task [30, 31]. Moreover, the NAO robot has been used previously as an entertaining interface for the maintaining of engagement during game-like activities for both children and adults [32]. However, a test interface, regardless of its engagement potential, is not entirely useful if it does not produce reliable data. As a first step towards the use of the robot as an auditory test interface, in this study, we will be investigating the reliability aspect of the data collected via the NAO robot, and in a separate study, the engagement based on human-robot interaction will be addressed [33].

In the field of social robotics, one of the most frequently used humanoid robots is the NAO from SoftBank Robotics. The use of the NAO has been suggested in the literature to facilitate testing procedures in hearing research [34–36]. Especially in human-robot interactions, the robot's relatively small size, its friendly and human-like appearance, and its sociable and non-judgemental characteristics seem to be helpful [37]. The NAO has been successfully used in previous studies as a therapeutic interface, motivating participants to learn and interact [38–42] (for an extensive review see [43]).

While the NAO could provide a good test interface for engagement purposes, the implementation of auditory and speech perception tests could be affected due to potential inherent limitations of the robot, such as sound quality (due to the internal speaker and sound card combination), non-experimental sound artefacts (due to the cooling fan and moving actuators [44]), stimulus processing speed (for tests that require stimuli to be prepared and processed in real-time during testing), and stimulus presentation and response logging [limited visual, voice and speech cues due to the non-moving face of the robot and the number of sensors (11 in total)]. On the point of sound quality, literature has also shown that the loudness of sound can be perceived differently depending on the source of the stimulus; e.g., from different loudspeakers and headphones, distance to the loudspeaker, hardware differences between output devices etc., despite the careful equalising between devices [45]. Furthermore, potential perceptual biases could be an additional factor such as the robot's voice being perceived as more of a specific gender if the physical characteristics of the robot are visually perceived as that specific gender [13]. Therefore, such an interface first needs to be confirmed to reliably produce good quality results for hearing and speech perception tests.

These experiments aim to evaluate how well the NAO would function as an auditory psychophysical testing interface, given the potential limitations and differences in implementation compared to the computer version, using tests of voice cue perception and subjective categorisation of voice gender.

## General methods

The present study is part of the larger project Perception of Indexical Cues in Kids and Adults (PICKA). The PICKA test battery was created by the dbSPL (for more details of the dbSPL group see www.dbspl.nl) research group at the University Medical Centre Groningen (UMCG) to investigate voice and speech perception in normal and impaired hearing. In addition to being part of the larger PICKA project, this study is also part of a larger study comparing the results of the four PICKA tests on the laptop to a humanoid NAO robot we named "Sam", chosen to represent a gender-neutral name in an attempt to avoid a prior gender assignment for the robot. The four PICKA tests are voice cue sensitivity, voice gender categorisation, voice emotion identification, and speech-on-speech perception. Two of these tests were used in this study, conducted as two experiments performed one after the other in a single session: Experiment I, voice cue sensitivity (similar to [8, 10–12]) and Experiment II, voice gender categorisation (similar to [7, 12]). The PICKA tests can be run both in English, Dutch and Turkish, the former of which was used in this and the larger comparative study. Until this study, the tests (developed in Matlab [46]) had been implemented on a laptop, some of which have interfaces with cartoons and animations, which we used in this study for a fairer comparison to the robot [12]. In each experiment, tests were performed both via the laptop (identical to that reported by [12]) and the new robot interface, Sam.

### General NAO robot setup

Sam is a NAO V5 H25 humanoid robot developed by SoftBank Robotics. The body of Sam has an Atom Z530 1.6 GHz CPU processor, 1 GB RAM, 2 GB flash memory, an 8 GB micro SDHC card, 11 tactile sensors–three on the head, three on each hand and one on each foot–two cameras and four ultrasound sensors. Sam has 25 degrees of freedom, enabling it to perform movements and actions resembling that of a human.

The operating system on Sam is the NAOqi OS, based on Gentoo Linux created by the original developers, Aldebaran. A cross-platform NAOqi SDK (software development kit) framework can be installed onto a local computer to communicate with and control Sam. The programming languages that can be used to interact with NAO through the SDK are Python [47], C++ [48], and Java [49].

Since the current version of the PICKA test battery was developed and designed in Matlab, this was not compatible with Sam if it were to function as an independent interface. Therefore, the PICKA tests were rewritten into Python, which allowed all tests and stimuli to be stored and run directly on Sam. However, it should be noted that the processor of Sam (1.6 GHz) is slower than the laptop (2.5 GHz); thus, from the beginning of the experiments, it was known that the real-time local generation of stimuli would possibly result in longer durations of the tests.

### Experimental setup

The laptop used was an HP Notebook (Intel Core i5 $7^{th}$ gen) running Ubuntu 16.04. The PICKA test battery was run using MATLAB 2019b. Stimuli for all tests were played through the internal speakers and sound card of the laptop. Responses were logged using a connected mouse to the laptop. The game-like interface with which children were previously tested [12] was used. Although all other details of the implementation were also identical to the aforementioned study, the only exception was the use of English stimuli in the present study, differing from the use of Dutch by [12].

For the robot setup, the stimuli for all tests were played through the internal stereo loudspeakers located in Sam's head and using the onboard soundcard. In both tests the tactile sensors on Sam's hands and head were used to log responses.

### Participants

Thirty adults participated in both experiments; however, two participants were excluded from data analysis due to not meeting the inclusion criteria for normal hearing, and data were analysed from 28 participants (aged 19–38; 23.6 ± 4.9 years; participants were asked with which gender they identified, to which they could respond openly: 19 reported female, 9 reported male). Sample size was based on a rule of thumb for human-robot interaction studies in which it is recommended that a minimum of 25 participants are included per tested condition [50], and an extra five participants to account for potential exclusions. Participant recruitment was conducted between 02/2021–09/2021, and inclusion criteria were kept general to minimise any selection bias. All participants reported English as either their native or first additional language and having completed at least high school education. Informed consent was obtained prior to the start of the experiment, followed by a pure-tone audiogram to confirm normal hearing for inclusion/exclusion [hearing thresholds > 20 dB HL at any of the audiometric octave frequencies (250–8000 Hz) qualified for exclusion]. Regardless of the outcome of the audiogram, the experiment was still conducted, and the inclusion/exclusion was applied before the data analysis phase. Although this deviates from common practice for psychophysical tests, an additional component of this study was to investigate the observed human-robot interaction, which will be reported in a follow-up publication. The PICKA project protocol was approved by the METc ethical review committee at the local university hospital (METc 2018/427, ABR nr NL66549.042.18). Participants provided written consent for their participation and were assigned a unique participant identifier, and the corresponding key was securely stored; however, the authors did have access to this information as necessary. The participants were compensated €8/hr for their participation.

### General setup

The order of the interfaces (i.e., starting with the laptop or Sam) in each experiment per participant was randomised. In a session, a break was offered to participants both between the two experiments and between the two interfaces within an experiment. On both interfaces and for both experiments in each session, a training phase (shorter version of the test) was first performed to familiarise the participant with how the test was conducted and how their responses were logged. After this, participants started with the actual test. During each experiment, participants' responses were recorded to assess performance for the specific auditory test with each interface.

In each experiment for each test, participants were seated at a desk with either the laptop or Sam in an unoccupied and quiet room at the university medical centre. Participants were seated approximately one metre from the test interface; however, this varied as participants moved to interact with Sam or the laptop. The unused interface was placed outside the participants' line of sight.

### Experiment I: Voice cue sensitivity

The voice cue sensitivity test assesses the listener's ability to detect the smallest perceivable F0 or VTL differences (just noticeable differences; JNDs) when applied to a speaker's voice (based on methods by [8]).

### Stimuli

To prepare the stimuli, consonant-vowel (CV) syllables were spliced from existing consonant-vowel-consonant meaningful English words from the Chear Auditory Perception Test (CAPT) and Consonant Confusion Test (CCT) corpora [51, 52]. The CV tokens had a duration of 142–200 ms. Splicing of syllables (60 in total) was performed identically to methods reported by [12]. For each trial, three spliced syllables were randomly selected and concatenated to produce a single CVCVCV syllable triplet (e.g., "bi-fo-ki").

For this test, the focus is on the difference in F0 and VTL relative to a reference voice. Both the F0 and VTL differences are expressed in semitones (st), an intuitive frequency increment unit often used in music and expressed as 1/12th of an octave. The VTL is a distance, hence related to wavelength and inversely to frequency, and can be expressed as ratios measured on a logarithmic scale ($12\log_2(r)$, where $r$ is the expansion/contraction ratio of the formant distances [8]). This conversion of the voice cues to semitone units allows both F0 and VTL values to be expressed in comparable units, instead of relying on the original Hertz or millimetre for F0 and VTL, respectively. To obtain the F0 contour and spectral envelope of each syllable when using the laptop, the analysis module STRAIGHT [53] was used. Extraction of these same parameters in Sam's Python implementation of the PICKA battery was performed using the analysis module PyWORLD [54] in place of STRAIGHT. Application of the modified voice cue parameters was made with methods identical to [8]. The F0 of the reference voice was set to 242 Hz. This reference F0 was the same value used by [12], despite using a different female speaker since the language was different (average F0 across all English syllable stimuli used was 248 Hz). This was done to make the results more comparable across studies. VTL is related to the distribution of the formant frequencies resulting from vocal tract resonances. Shortening of the VTL by the expansion/contraction ratio of the formant distance shifts all the formants to a higher frequency by that same ratio. Therefore, a positive VTL change corresponds to a negative formant frequency shift in semitones [4]. The VTL of the reference voice was left unchanged from the original speaker.

All stimuli for both interfaces were calibrated to 65 dB SPL using a Knowles Electronics Mannequin for Acoustic Research (KEMAR, GRAS, Holte, Denmark) head assembly and a Svantek sound-pressure level metre (Type 2610, Brüel Kjær and Sound & Vibration Analyser, Svan 979).

### Laptop vs robot

**Interface.** The laptop game-like interface of the voice cue sensitivity test can be seen in Fig 1, panel A. The three-syllable triplets were presented by each of the three identical aquatic animals. The participant then clicked the animal that sounded different from the other two, and visual feedback was given for correct responses by fish and sea creatures moving. Sam as a

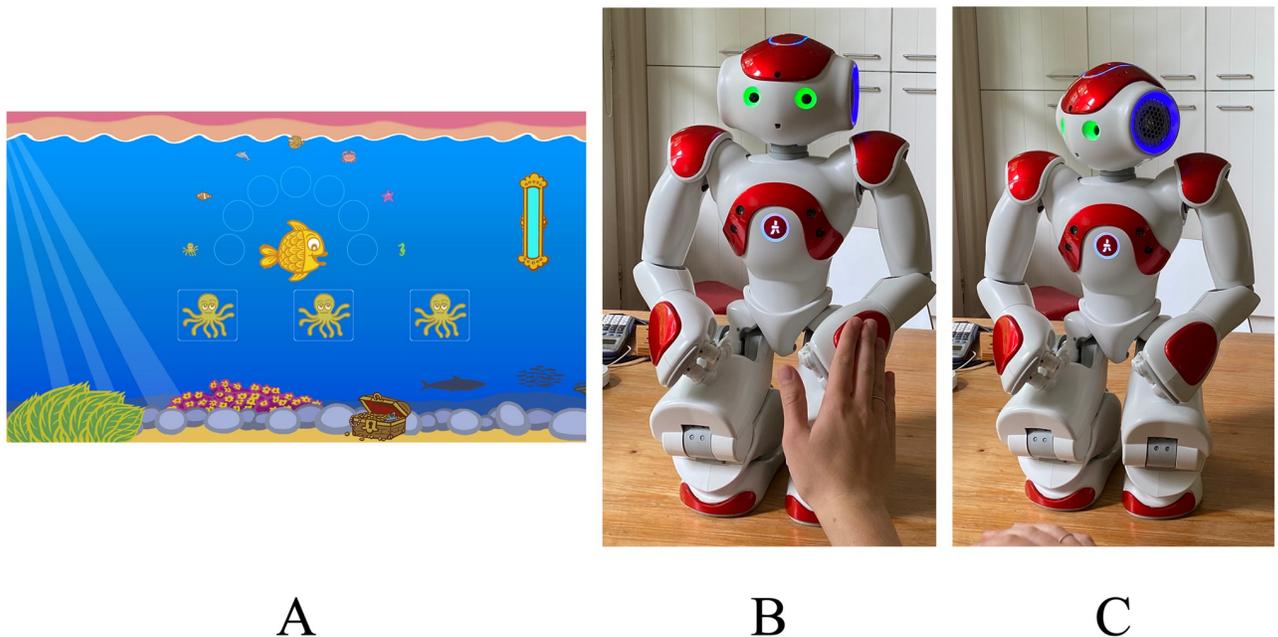

**Fig 1. Voice cue sensitivity test as presented via the laptop game-like interface (A) and Sam (B, C).** In (A), the three octopi produce voice stimuli, and the participant enters the different voice by clicking on the corresponding octopus. Positive feedback is given by the fish doing a circular-dance and the octopus swimming towards and joining the other sea creatures. Feedback was presented by the outlining of the correct response in the case of incorrect answers and; the next stimulus was played after logging of an incorrect response. The sand clock in the top right indicates progress proportional to the number of trials completed. The illustrations were made by Jop Luberti for the purpose of the PICKA project. This image is published under the CC BY 4.0 licence (https://creativecommons.org/licenses/by/4.0/). In (B), a response is logged by touching Sam's left hand. The green eyes indicate that a response can be given. In (C), visual feedback in the form of head movements is presented indicating if the response was correct or incorrect. With Sam there is no indicator of progress.

https://doi.org/10.1371/journal.pone.0294328.g001

test interface can be seen in Fig 1, panels B and C. Tactile sensors on Sam's hands and head (the three sensors on each hand and the head were grouped together as one) were used for response logging, and visual feedback was given for correct responses by Sam nodding.

**Implementation.** Implementation of the test on both interfaces was fundamentally similar regarding the on-the-fly preparation of stimuli, adaptive procedure, conditions for terminating blocks, visual feedback during the testing and training phases, and presentation of the stimuli from the interface itself (i.e., no external sound cards or speakers were used). Differences in the implementation concerned the human robot interaction. This included Sam introducing themself and the PICKA test to the participant, and Sam's verbal encouragement and offer of a break between blocks, detailed below.

**Procedure.** In each trial, the listener was presented with three acoustic stimuli, each made up of the same triplet of different syllables, where one of the three stimuli differed either in F0 or in VTL relative to the other two reference triplets. The task for the listener was to identify which of the three presented stimuli sounded different from the other two. The overall paradigm followed a three-interval three-alternative forced-choice (3I-3AFC) adaptive 2-down-1-up staircase model, converging to 70.7% correct discrimination [55]. After two consecutive correct responses, the relative difference between the different and the reference stimuli was reduced, making identification of the differing stimulus more difficult. After one incorrect response, the relative difference increased, making identification of the different stimulus easier. The test comprised four runs representing four directions of voice manipulation: two for F0 discrimination and two for VTL discrimination. Each run started with a difference of either -12 st or + 5 st for ΔF0 (corresponding to values typical of male and child talkers, respectively) or +3.8 or -7.0 st for ΔVTL (corresponding to values typical of male and child talkers, respectively). From this starting point, the voice cue difference decreased (i.e., approached 0 st difference compared to the unmodified reference voice) after two consecutive correct responses or increased after one incorrect response with a predetermined step size, initially set at 2 st. As more correct responses were given, the step size was also adapted by reducing the previous step size by a factor of $\sqrt{2}$, becoming exponentially smaller such that the step size approached but never reached a 0 st voice cue difference. Although the voice cue difference increased after an incorrect response, the step size was not modified. A reversal was defined when a single incorrect response was given after at least two correct responses, or two correct responses were given after at least one incorrect response. Each run ended in one of three ways: 1) if 15 consecutive incorrect responses were given, 2) if a total of 150 stimuli were presented or 3) after eight reversals had been reached. When a run ended in the latter, the JND was calculated by averaging the difference in semitones over the last six reversals. The former two conditions were implemented as a measure to ensure the test would not continue indefinitely, or in case the participant could not continue with the test. However, this did not occur in any of our experiments; all participants finished each run after the eight reversals.

Participants first familiarised themselves with the test through a training consisting of six randomly selected practice stimuli, identical to test stimuli but performed with a larger, fixed step size of 3 st to speed up the adaptive procedure. The syllable triplets used in the training stimuli were not reused in the testing phase. Following the training, the first of the four test runs was started, the order of which (voice cue and direction) was randomised for each participant. In both the training and testing phases on both interfaces, positive visual feedback was given to participants. On the laptop, this was provided by the central fish turning in a circle, and the aquatic animal representing the correct response "swimming" to a growing line of sea animals representing previous correct responses. Although no explicit negative feedback was presented for incorrect responses, the correct response is briefly outlined in green before continuing with the next stimulus without any further animations. At the end of each run, a "Start" button was displayed to begin the next run, and the participant was allowed to take a short break before starting the next run; however, this was not explicitly indicated to participants.

On the laptop, the three aquatic animals presented the stimuli; on Sam, specific tactile sensors corresponded to the order in which the stimuli were presented: the first stimulus corresponded to Sam's right hand, the second to the head and the third to Sam's left hand. After each stimulus, Sam's eyes changed colour from white to green to indicate that a response could be given, after which the eyes returned to white. This was implemented to prevent participants from logging their responses too early before the stimulus had finished playing. After each response, visual feedback was presented as either a head nod if correct or a head shake if incorrect (Fig 1, panel C). Although the addition of the explicit negative feedback differs from the implementation of the laptop, the lack of a screen with the robot, unlike the laptop, may make it unclear as to whether or not a participant's response was logged without such an explicit visual cue. To maintain engagement throughout each run, Sam autonomously encouraged participants to continue depending on their performance. The choice of whether or not to provide encouragement was decided by a randomly generated number between 0 and 1. If the number was less than some threshold, initially set at 0.1, encouragement would be provided. If the previous response was correct, Sam would say either "Keep going!" or "Doing well.". If the previous response was incorrect, Sam would motivate them by saying either "Give it another go" or "Keep trying". Every time the response was incorrect, the encouragement threshold was increased by 0.05 until encouragement was given, after which the threshold was reset to 0.1.

After each run, Sam asked the participant if they wanted to take a break, to which they could verbally respond with either "yes" or "no". If they responded "yes", Sam would ask if they would like to stand up and follow along in a stretch routine, to which they could again verbally respond. If the participant chose not to take a break, the next run would start.

## Data analysis

To determine if the two implementations were comparable for the voice cue sensitivity task performance, JND thresholds from either interface were first log-transformed to convert the data into a normal distribution, as the thresholds are always a positive value. Repeated-measures ANOVAs were performed for each voice cue, F0 and VTL, separately using the interface the test was performed on, and the direction (negative or positive) of the vocal cue as the two within-subjects factors (two interfaces ⨯ two voice cue directions). To improve the robustness of the data analysis, Bayesian repeated-measures ANOVAs were performed for each voice cue using the same within subject factors mentioned above. Similar classical and Bayesian repeated measures ANOVAs were performed to examine the effect of the interface and cue direction on the duration of each test run.

Bayesian inferences were used in this study because we are looking for evidence that the two interfaces are similar, and this type of conclusion cannot be reached with a classical (frequentist) approach. In the frequentist approach, the *p*-value is the probability of obtaining results at least as extreme as those seen in the collected data given that the null hypothesis ($H_0$) was true [56]. Therefore, a lack of significance is often falsely interpreted as the absence of an effect [57], which we may be tempted to interpret as the two interfaces being equivalent. In comparison, a Bayesian analysis allows for an alternative interpretation and reasoning of the results through the reporting of the magnitude of evidence [i.e., the likelihood of the data under the assumption of $H_0$ rather than H1 (the alternative hypothesis)]. An estimate of this evidence is presented as the Bayes' factor, which provides the relative likelihood of the data with respect to the null hypothesis ($BF_{01}$ = H0/H1) or any other hypothesis ($BF_{10}$ = H1/H0). it

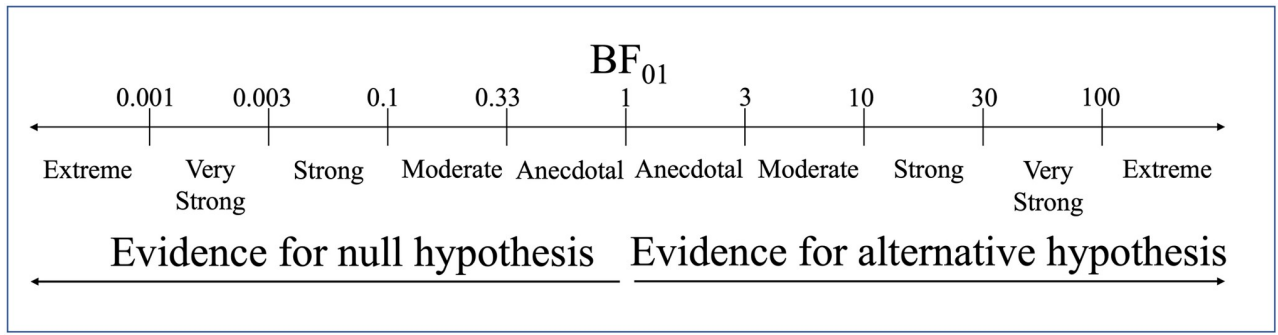

**Fig 2. Classification scheme for the Bayes factor (BF01) by JASP.**

https://doi.org/10.1371/journal.pone.0294328.g002

should also be noted that these two notations of the Bayes' factor are indeed reciprocals of one another (i.e., $BF_{01} = 1/BF_{10}$). Bayesian analyses were performed using the statistical software JASP [58], which categorises the evidence based on the Bayes' factor as shown in Fig 2. Further discussion on classical vs Bayesian inference is presented in the General Discussion (also see [56] for a detailed introduction to the Bayesian method).

Organisation of the data was performed using the data analysis software R [59], and JASP was used for all statistical analyses. Bayesian analyses for both vocal cues used the default priors suggested by JASP: a uniform model prior with *r scale fixed effects* = 0.5, *r scale random effects* = 1 and *r scale covariates* = 0.354 and enforced principle of marginality on the fixed effects. All processed data is openly available at [60].

## Results

Fig 3 depicts JND thresholds obtained with each interface and previously reported thresholds in previous studies [6, 8]. Laptop F0 JNDs [2.453 ± (standard deviation of) 2.130] were on average 1.212 st larger than Sam's F0 JNDs (1.241 ± 1.146), and laptop VTL JNDs (1.828 ± 1.124) were on average 0.450 st larger than Sam's VTL JNDs (1.378 ± 0.725). Results of the repeated measures ANOVA (main factor of test interface with two levels; laptop, robot) showed no statistically significant difference between the JND thresholds for F0 due to the interface [$F(1, 28) = 4.170, p = 0.051, \eta^2_p = 0.130$], nor was there a cue direction effect [$F(1,28) = 0.035, p = 0.852, \eta^2_p = 0.130$]. Similarly, there was no significant effect of the interface on VTL vocal cue [$F(1,28) = 0.032, p = 0.859, \eta^2_p = 0.001$]; however, there was a significant effect due to cue direction [$F(1,28) = 5.337, p = 0.028, \eta^2_p = 0.160$]. Results showed no significant effect for an interaction between the interface and the direction of the F0 cue [$F(1,28) = 0.005, p = 0.946, \eta^2_p = 1.65e-4$], but indeed a significant effect for an interaction between the interface and VTL voice cue direction [$F(1,28) = 5.578, p = 0.025, \eta^2_p = 0.166$]. However, following a post-hoc Bonferroni-Holm correction test, the significant effect was only between the directions of the VTL cues on the laptop, not between the two interfaces.

The average duration of the voice cue sensitivity test (including all 4 runs) was 19 ± 2.7 min on the laptop and 29 ± 3 min on Sam. These times are exclusive of breaks taken by participants. On average, most participants did not take a break when using the laptop, whereas an additional two minutes on average was due to breaks when using Sam. Repeated measures ANOVA (main effect of test interface with two factors: laptop, robot) showed that Sam took significantly longer to complete in comparison to the laptop for the F0 vocal cue [$F(1,28) = 52.964, p < 0.001, \eta^2_p = 0.654$], but cue direction had no effect [$F(1,28) = 1.472, p = 0.235, \eta^2_p$

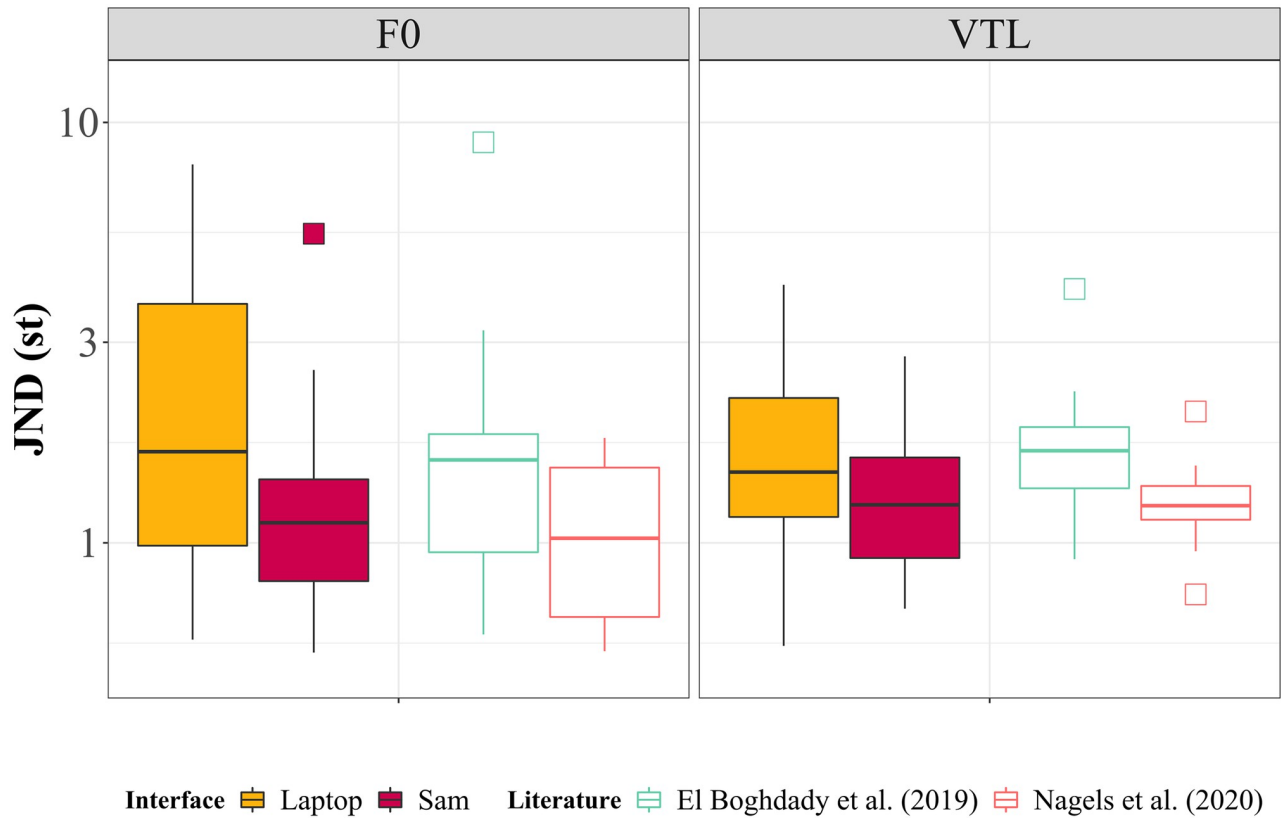

**Fig 3. Comparison of just-noticeable difference (JND) thresholds from the laptop, Sam, and previous studies, (left to right in each panel) shown for F0 and VTL in left and right panels, respectively.** Boxes indicate the range, quartiles, and median thresholds. Lower thresholds indicate better performance in the voice cue sensitivity test. Yellow = Laptop, maroon = Sam, cyan = normal hearing adult data from [11], red = normal hearing adult data from [12].

https://doi.org/10.1371/journal.pone.0294328.g003

= 0.050]. Similarly, the VTL vocal cue showed that Sam took significantly longer to complete the test [$F(1,28) = 52.670$, $p < 0.001$, $\eta^2_p = 0.653$], but no effect due to cue direction [$F(1,28) = 3.603$, $p = 0.068$, $\eta^2_p = 0.114$]. For both vocal cues, there was no effect on the duration due to an interaction between the interface and the direction of the cue [F0: $F(1,28) = 2.524$, $p = 0.123$, $\eta^2_p = 0.083$; VTL: $F(1,28) = 0.026$, $p = 0.873$, $\eta^2_p = 9.232\text{e-}4$].

Bayesian repeated measures ANOVAs showed strong evidence that Sam took longer to complete the test for the F0 vocal cue ($BF_{10} = 2.133\text{e+}5$), and anecdotal evidence that cue direction did not affect the duration of the F0 test runs ($BF_{10} = 0.388$). Bayesian results also showed that Sam took longer to complete the test for VTL vocal ($BF_{10} = 1.652\text{e+}5$), and anecdotal evidence that the cue direction also affected the duration ($BF_{10} = 1.035$). Furthermore, there was anecdotal evidence of an interaction between the interface and cue direction for the F0 vocal cue ($BF_{10} = 1.120$), and moderate evidence of no interaction for the VTL vocal cue ($BF_{10} = 0.261$). Fig 4 depicts the durations to complete each of the four test runs in the present experiment, as well as a comparison to data reported by [12]. Data reported by [11] is excluded from the analysis of test duration as they had used a different interface for the test (i.e., they did not use the same game-like interface as used in this experiment or by [12]).

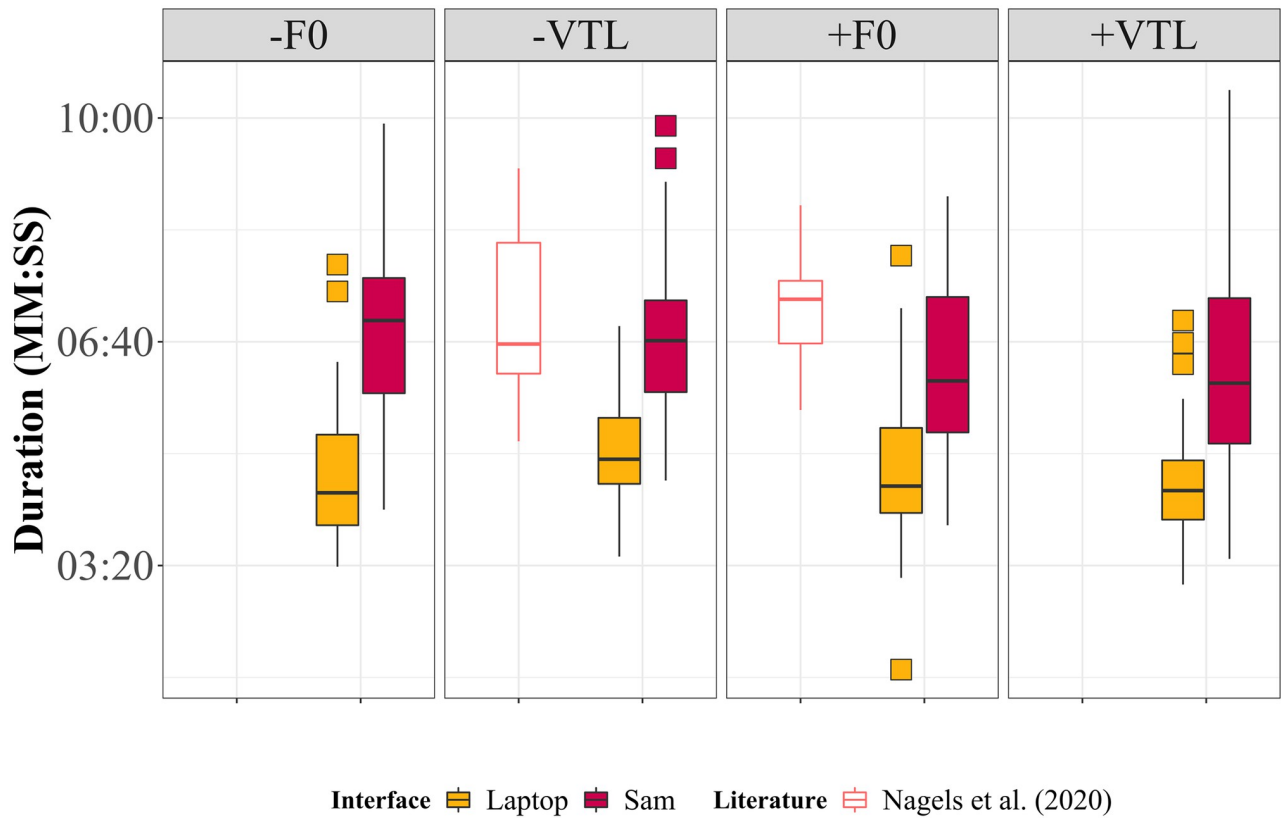

**Fig 4. Box plot depicting the duration to complete the four test runs of the voice cue perception task, as well as a comparison to data reported by [12].** Data reported by [12] only considered two of the test conditions: -VTL and +F0.

https://doi.org/10.1371/journal.pone.0294328.g004

## Discussion

In Experiment I, we implemented a test of voice cue sensitivity on Sam to compare JND measurements with the current laptop game-like interface in order to evaluate the viability of using a NAO robot interface for auditory psychophysical tests. Results for the voice cue sensitivity test showed no statistically significant difference in JND thresholds between the two interfaces concerning both F0 and VTL. For both voice cues, the thresholds on Sam were overall either smaller or similar to those obtained on the laptop, meaning that there was no indication for the performance being worse or the responses provided being less accurate when using Sam. This is a confirmation that whatever technical limitations Sam may have that could impact sound quality or sound processing, did not negatively affect the JND thresholds for this test with the young adult, normal-hearing population tested, compared to a laptop implementation.

Interpretation of the Bayesian analysis results is presented as follows. Bayesian analyses showed that F0 JND thresholds were 1.6 times more likely to result in a non-zero difference between the means; however, this evidence is only anecdotal; any variation in the means between the two interfaces is likely due to the sample size. In comparison, Bayesian analyses showed that the VTL JND thresholds were 3.3 (reciprocal of $BF_{10}$ = 0.304) times more likely to result in a zero difference between the means. Regarding the direction of the voice cue, Bayesian analyses showed that for F0, results were 5.3 times more likely to have no effect on the overall JND threshold, and 1.1 times more likely to affect the VTL JND threshold. With respect to interactions between the direction of the voice cue and the interface, the interaction was 3.8 times more likely to have no effect on the F0 JND threshold, but 4.6 times more likely to affect the VTL JND threshold.

In addition to comparing the results of the voice cue sensitivity test between the interfaces, when comparing the current results to those of previous studies using similar procedures, it was also observed that the F0 JNDs on Sam when using English stimuli were not significantly different from those reported by [12] [$t(51.45) = 1.33$, $p = 0.189$, Cohen's d = 0.322; $BF_{10}$ = 0.39] nor from those reported by [11] [$t(34.795) = -1.404$, $p = 0.17$, Cohen's d = 0.500; $BF_{10}$ = 0.88], both of whom had used Dutch stimuli. In comparison, the F0 JNDs from the laptop were found to be statistically different from the data reported by [12] [$t(70.953) = 4.777$, $p < 0.001$, Cohen's d = 0.940; $BF_{10}$ = 5.82], but were not statistically different from the data reported by [11] [$t(49.096) = 0.910$, $p = 0.367$, Cohen's d = 0.260; $BF_{10}$ = 0.38]. A closer inspection of the data reveals that a few participants obtained much higher thresholds on the laptop in comparison to their matched conditions on Sam, causing the expanded threshold range seen in Fig 3. The three participants with the largest deviation between the two interfaces, listed in Table 1, started the experiment on Sam; thus, it is unlikely that a learning effect could have caused the observed variations. It is possible, however, that after performing the test on Sam, participants were fatigued and were not as attentive to the task when performing it for the second time on the laptop. Furthermore, because the starting interface was balanced across all participants, the absence of this asymmetry may suggest that using Sam is motivational enough to combat the effects of fatigue when performing the test for a second time with Sam.

Since the same stimuli were used on both interfaces, ideally, it is expected that any variation between the interfaces would be identical. However, based on the t-test and effect size comparing the JNDs obtained on the laptop with those obtained by [12], it is likely that the large discrepancies seen in Table 1 could be related to the quality of the speakers on either interface, as their respective built-in speakers were used, and not external higher quality speakers. In addition, the procedure used by [12] made use of headphones instead of the loudspeakers; however, speakers were used in this study for consistency with the speakers of Sam. Moreover, both studies by [11, 12] included native Dutch speaking participants, and thus could be more selective with their inclusion criteria. In comparison, also being in the Netherlands and conducting the tests in English, participants included in the present study were only required to have a good understanding of English and did not have to be native speakers. Although the speakers on the laptop may have been of poorer quality compared to what is typically used in auditory research, the consistency in results obtained on Sam with those obtained in previous

**Table 1. Comparison of largest participant threshold discrepancies between the laptop and Sam.** NHA = normal hearing adult.

|                |                 | JND Threshold (st) |      |
| :------------: | :-------------: | :----------------: | :--: |
| Participant ID | Voice Condition | Laptop             | Sam  |
| NHA013         | +F0             | 5.82               | 2.99 |
| NHA023         | +F0             | 6.45               | 1.05 |
|                | -F0             | 5.66               | 1.79 |
| NHA027         | +F0             | 8.42               | 1.09 |

Comparing the VTL thresholds obtained on Sam and the laptop with previously reported data showed no significant difference between Sam and data reported by [12] [$t(43.15) = 1.23$, $p = 0.23$, Cohen's d = 0.25; $BF_{10}$ = 0.39], but indeed a significant difference compared to [11] [$t(51.42) = -2.14$, $p = 0.037$, Cohen's d = -0.51; $BF_{10}$ = 2.03]. There was also a significant difference between VTL thresholds obtained on the laptop and [12] [$t(71.143) = 3.67$, $p < 0.001$, Cohen's d = 0.60; $BF_{10}$ = 1.66]; however, there was no significant difference with [11] [$t(80.152) = 0.59$, $p = 0.56$, Cohen's d = 0.11; $BF_{10}$ = 0.26].

https://doi.org/10.1371/journal.pone.0294328.t001

studies does show that the JND test is robust not only across languages, but also regarding implementations and procedures.

It was seen that the duration of the test on Sam was longer than on the laptop; however, the duration of the test did not differ significantly between Sam and from durations reported by [12] [F0: $t(39.48)$ = -1.206, $p$ = 0.235, Cohen's d = -0.336, $BF_{10}$ = 0.482; VTL: $t(30.294)$ = 0.794, $p$ = 0.433, Cohen's d = 0.246, $BF_{10}$ = 0.394]. A similar comparison regarding the duration could not be made with data reported by [11] as they had used a different interface than that used in this experiment (i.e. they did not use the game-like version of the PICKA test battery). To investigate the longer running time on Sam, we have also inspected the number of trials on each interface. On average across all runs, the number of trials to complete in a block was 37 ± 8 on the laptop, and 42 ± 8 on Sam. Although more trials were needed on average to achieve the JND threshold on Sam [$t(115)$ = 4.113, $p < 0.001$, Cohen's d = 0.561, $BF_{10}$ = 231.775], this difference does not fully explain the overall longer duration of the test. A perhaps more important factor that likely influenced the duration of the test was the processing time of the stimuli. The average processing time for a single stimulus was around two seconds on Sam, and one second on the laptop. This is a limitation of Sam's hardware, and it is expected that newer models of the NAO could solve this processing discrepancy. Despite the longer testing duration on Sam, both including and excluding the breaks, Sam was still able to collect comparable JND thresholds to the laptop interface.

One could argue that a more powerful humanoid robot, such as Pepper (another of Aldebaran's robots), could be used in place of Sam to compensate for the processing delays. While this is technically true as both robots use (nearly) identical software platforms, are designed to portray a friendly agent, and have been used in various HRI scenarios such as engagement and social robot application (for a review please see [37]), Sam offers other advantages over Pepper for our specific application, most notably, its small size and lower cost. Sam's smaller size makes it easily transportable to various testing locations. In addition, should anything go awry, a smaller robot minimises the risk of potential harm to users, especially in smaller testing locations such as clinical rooms. A simpler solution to the low processing power of Sam could be the offline generation of all JND stimuli through simulations, and then use the pre-generated stimuli during testing. However, this could introduce a new problem of storage space on Sam. The incorporation of a stronger processor that could be attached to Sam as a type of "backpack" [61] could be another solution to the processing capacity of Sam. Instead of replacing Sam's central processing unit (CPU), an additional CPU could be used in parallel to perform the more process intensive tasks. There are a number of ways in which processing time for the JND stimuli could be reduced, all of which could be thoroughly explored as future improvements.

It should also be considered that the inclusion of a dedicated break offered to participants after each test block could potentially have an effect on the performance results. It has been shown in literature that the duration one interacts with an agent, the perception they have towards the agent may change [62, 63]. [62] has shown how over the course of an interaction, perceptions of warmth and competence of a virtual agent can change, whereas [63] has shown how during an interaction with a robot, positive interactions tend to result in more favourable perceptions of the robot. Following this, there is a possibility that the inclusion of the follow-along stretch routine during the break may have increased the likeability of Sam, potentially motivating participants to exert more effort in proceeding test blocks.

### Experiment II: Voice gender categorisation

The voice gender categorisation test investigates the subjective categorisation of a speaker's expressed gender in their voice, and how voice cues and manipulations can influence the perception of voice gender. This test is based on the methods by [7, 12].

### Stimuli

To investigate this voice cue influence, pre-generated stimuli from English consonant-vowel-consonant spoken words taken from the same corpora as in Experiment I (CAPT and CCT) were used. Each stimulus presented during the test was randomly chosen from a limited list of words: "bike", "pool", "watch" and "hat"; which were altered in F0 and VTL. Similar to methods described by [12], stimuli were root-mean equalised, after which stimuli were modified using the STRAIGHT Matlab module on the laptop or PyWORLD on Sam. F0 and VTL were manipulated independently from each other, and in cases where both cues were altered in a single stimulus, F0 alterations were first applied followed by VTL alterations. When no alterations were made (F0 = 0.0 st and VTL = 0.0 st), the original stimulus was still resynthesized to account for any potential synthesis artefacts. Three levels of modifications were used for each voice cue in all stimuli, identical to that used by [12]; decreasing F0 from 0.0 to -6.0 st and -12.0 st and increasing VTL from 0.0 to +1.8 st and +3.6 st. In addition, combinations of these voice cues were used for each word. This resulted in nine voice alterations for each word, producing 36 pre-processed stimuli (nine voice conditions ⨉ four words) for the block. To familiarise participants with the test, the same eight example words were used for all participants, taken from the 36 stimuli at the two widest vocal manipulation conditions (four words ⨉ two widest voice conditions–F0 = 0.0 st, VTL = 0.0 st, and F0 = -12.0 st, VTL = +3.6 st).

### Laptop vs. robot

**Interface.** For the voice gender categorisation test on the laptop's game-like interface, when the stimulus was presented, a cartoon image of either a male or female was shown on an animated television screen. Participants had to agree or disagree if the presented voice gender matched the image gender. When performing the test on Sam, only the tactile sensors on the hands (the three sensors on each hand grouped together as one) were used to log responses.

**Implementation.** Similar to Experiment I, the core paradigm of the voice gender categorisation test was similar between the two interfaces, and stimuli were presented without the use of external sound cards or speakers. Sam provided an introduction and explanation of the test before the training phase, similar to Experiment I. Unlike Experiment I, however, stimuli were pre-processed for both interfaces. Additionally, no visual feedback was provided to participants from either interface, as the voice gender categorisation was a subjective choice; nor was there any encouragement provided or any breaks offered due to the shorter duration of the test in comparison to Experiment I.

**Procedure.** The test consisted of one block in which all 36 aforementioned stimuli were presented in a randomised order. Each stimulus was presented once, after which the listener had to decide on whether the voice sounded male or female. While a person's expressed gender is more flexible and wider than these two categories [64], for methodological simplicity and to enable a comparison to previous work, we have followed the previous procedures [7, 12]; and as a result, participants were only given the options of these two categories.

Participants were first presented with a training phase consisting of the eight example words to familiarise themselves with the task procedure. Thereafter, the data collection was started. An example of the test interface on the laptop is shown in Fig 5. In both training and data collection, there was no visual feedback provided, as the voice gender categorisation is a subjective choice; and, therefore, there was no (in)correct categorisation. Since the test consisted of one block, only a break was offered between the two interfaces.

The laptop game-like version required participants to click on either a green tick to indicate that the presented voice gender and the picture gender matched, or a red cross if they did not match. In comparison, Sam's hands corresponded with either a male or female categorisation,

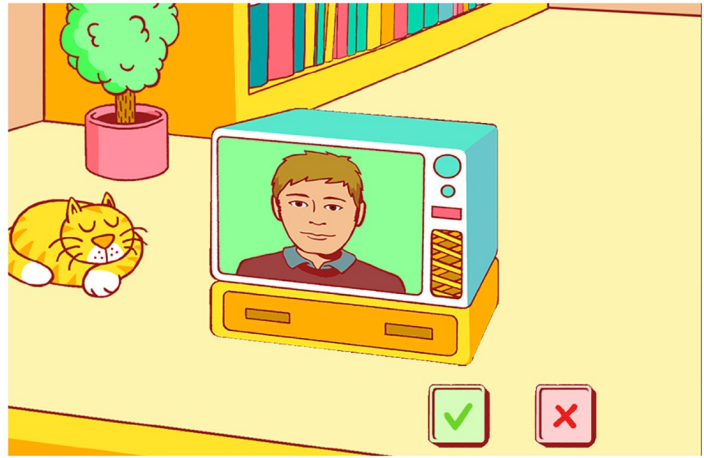

**Fig 5. Voice gender categorisation as presented on the laptop.** After each stimulus was presented, the image on the TV would change to show either a male or female person and participants would either agree (clicking the green tick) if the voice matched the new image or disagree (clicking the red cross) if the two did not match in their subjective opinion. The illustrations were made by Jop Luberti for the purpose of the PICKA project. This image is published under the CC BY 4.0 license (https://creativecommons.org/licenses/by/4.0/).

https://doi.org/10.1371/journal.pone.0294328.g005

randomising the hand-gender pair after each stimulus was presented. The purpose of such randomisation was to avoid a bias for participants continuously touching the same hand, whilst bringing their attention back to Sam. After the randomisation, Sam indicated which hand corresponded with which gender before allowing the participant to log their response. This was done by Sam lifting and rotating its hand outward before returning it to its default position on its legs. Although this differs from the laptop implementation, the latter does also include a degree of randomisation when the image of a male or female person is randomly presented. An increase in the test duration as a result of this method of indicating the gender-hand pair was taken into account; however, based on the relatively short duration of this test from the literature, it was presumed this would not heavily affect the average duration of the test. After Sam indicated the gender-hand pair, visual cues were provided by changing the colour of Sam's eyes (similar to Experiment I): from white to green indicated a response could be given, back to white indicated the response had been logged.

### Data analysis

Analysis of the categorisations made during the voice gender perception test was performed similarly to that carried out by [12], whereby cue weights were calculated as a perceptual weighting of F0 and VTL for participants' categorisation judgments, effectively splitting categorisations as a function of the two voice cues. Calculations were made by first normalising F0 and VTL relative to the reference speaker's voice and were defined as follows: $\delta F0 = -\Delta F0/12 - 0.5$, $\delta VTL = \Delta VTL/3.6 - 0.5$, thus making the two voice cues functionally equivalent. This resulted in the reference female voice, which had an $\Delta F0 = 0.0$ st, a $\Delta VTL = 0.0$ st, corresponding to a $\delta F0 = -0.5$ and a $\delta VTL = -0.5$; and the male-sounding voice with an $\Delta F0 = -12.0$ st, $\Delta VTL = +3.6$ st, corresponding to a $\delta F0 = +0.5$ and a $\delta VTL = +0.5$. Using a mixed-effects logistic regression model with slopes for $\delta F0$ and $\delta VTL$ per participant, the coefficients for each participant could be extracted, in lme syntax: response ~ ($\delta F0 + \delta VTL$|participant). These coefficients provide a prediction on a logit scale relative to the normalised $\delta F0$ and $\delta VTL$ ranges, which can subsequently be converted into "Berkson" units (Bk) per semitone so that they correspond to a $\log_2$ odds ratio per semitone [12, 65]. Using the calculated Bk units and cue weights (Bk units per semitone difference) of the models. Paired samples student t-tests were performed to test for differences between the two interfaces. As with Experiment I above, Bayesian paired samples t-tests were also performed to improve the robustness of the statistical results. Bayesian analyses were carried out using the default Cauchy prior (? = 0.707) and a null hypothesis that the means between the two interfaces were equal.

### Results

Fig 6, panel A depicts the distribution of the cue weights for each interface and panel B the contribution of F0 and VTL on the voice gender categorisation. The colour of each square indicates the frequency of categorisation based on the influence of VTL difference (dVTL) and F0 difference (dF0). Squares coloured toward the yellow side of the spectrum indicate more

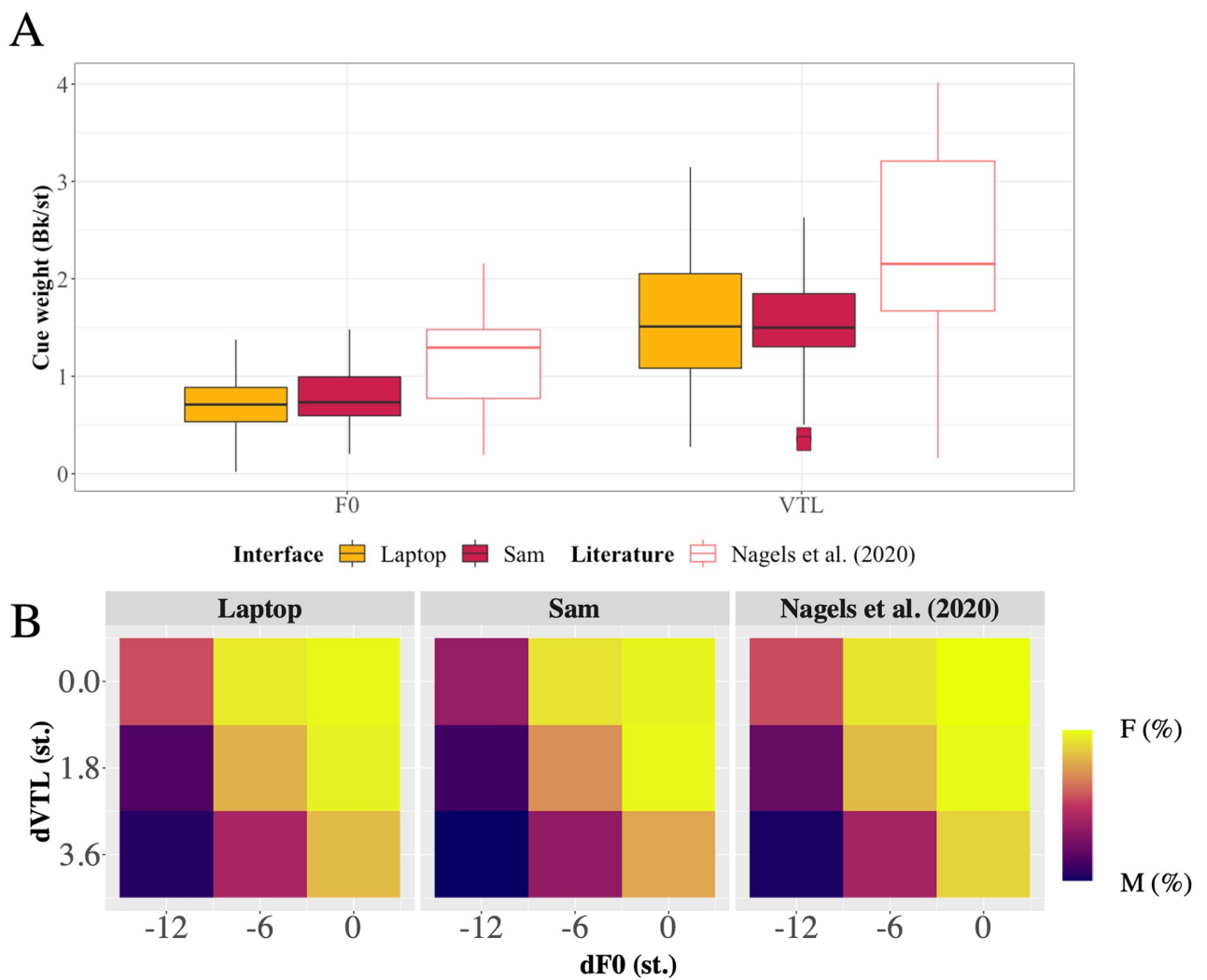

**Fig 6.** Panel (A) is a boxplot depicting the overall cue weightings (Bk/st) of F0 and VTL for voice gender categorisation, shown separately for the two interfaces and in comparison, to previously reported data by [12]. Boxes show quartiles, and the line in each box shows the median cue weights. The dots indicate outliers' cue weights. Panel (B) is a mapping of the voice gender categorisation judgments as a function of the contributions of ΔF0 and ΔVTL voice cue differences (dF0 and dVTL, respectively) on voice gender categorisation for each interface, and a comparison to previous data from [12].

https://doi.org/10.1371/journal.pone.0294328.g006

female categorisations, whilst squares toward the dark blue side indicate more male categorisations. Results of the paired student t-tests showed a statistically significant difference between the means of the intercepts [$t(28) = 2.549$, $p = 0.017$, Cohen's d = 0.455], but no significant difference between the means of the cue weights for the F0 [$t(28) = -0.952$, $p = 0.349$. Cohen's d = -0.177] and VTL [$t(28) = 0.53$, $p = 0.6$, Cohen's d = 0.135] voice cues.

Bayesian analyses between the two interfaces showed strong evidence that the two interfaces were different with regards to the means for the intercepts ($BF_{10} = 43.21$), but no difference between the means for the cue weights of the F0 model ($BF_{10} = 0.298$), or the cue weights of the VTL model ($BF_{10} = 0.225$). Bayesian results also showed anecdotal evidence for a difference in gender categorisation between the two interfaces ($BF_{10} = 1.248$).

The average duration to complete the test was 3 min ± 17 s on the laptop, and 5 min ± 49 sec on Sam. Comparison of the duration to complete the test between the two interfaces showed that Sam took significantly longer to complete [$t(28) = 15.840$, $p < 0.001$, Cohen's d = 3.717; $BF10 = 3.567e+12$]. This difference in duration can be seen Fig 7 below, also comparing the durations to that reported by [12].

## Discussion

In Experiment II, we implemented a voice gender categorisation task on Sam to compare the categorisation of a voice's gender between Sam and the currently used laptop interface.

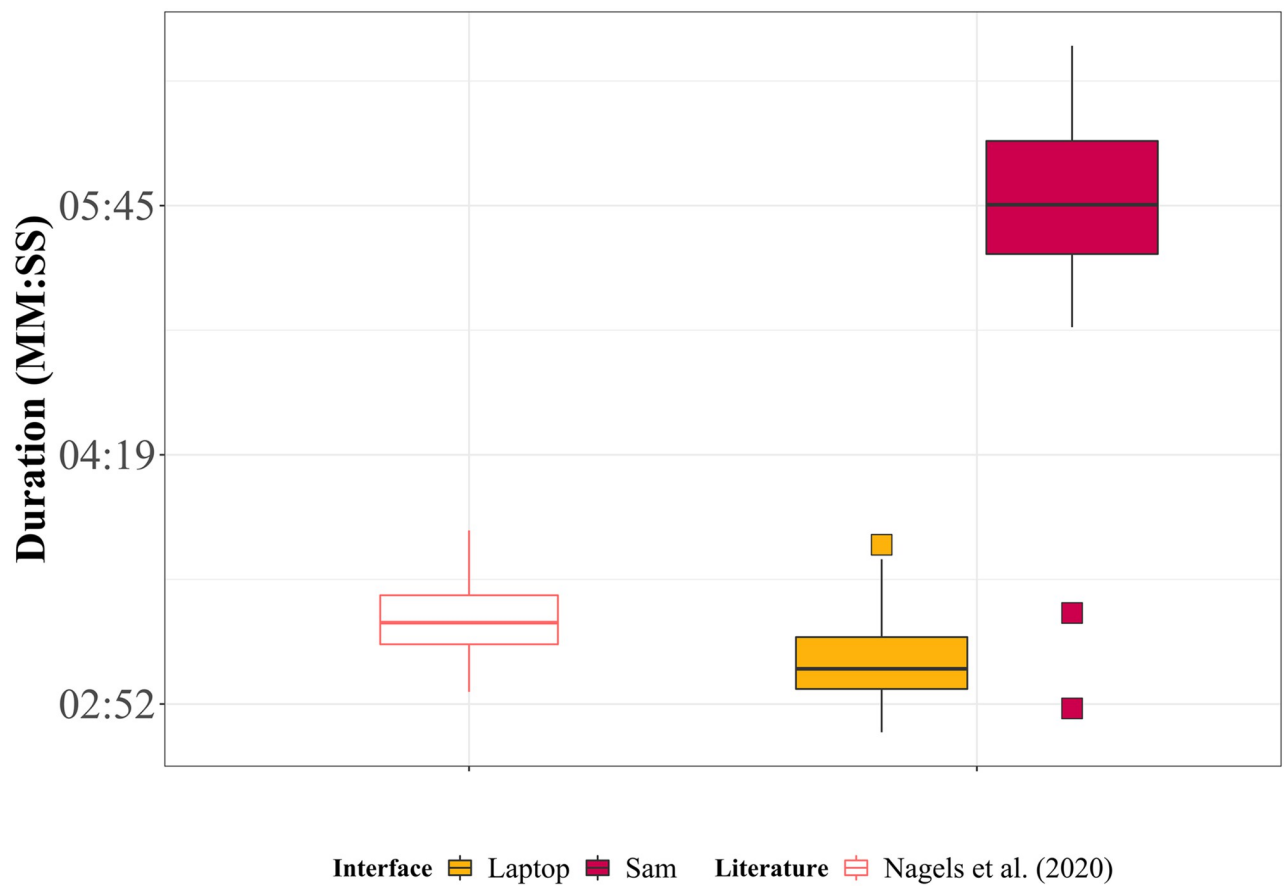

**Fig 7. Boxplots depicting the duration to complete the voice gender categorisation task on the laptop and Sam interfaces, as well as a comparison to data reported by** [12].

https://doi.org/10.1371/journal.pone.0294328.g007

Classical analyses showed a significant difference between the gender categorisation between interfaces and intercepts of the cue weights, but no difference between the cue weights themselves. Bayesian analyses of the Bk units between the two interfaces showed strong evidence (i.e., 43.2 times more likely) for a difference in the cue weight intercepts. Cue weights for F0 and VTL also showed moderate evidence (4.35 and 3.33 times more likely, respectively) of no difference between the two interfaces. Bayesian analyses showed anecdotal evidence for a difference in overall categorisations of gender between the two interfaces; however, the use of the Bayesian method here indicates that the significant difference seen in the classical method indeed has a small effect size, and thus any differences observed are likely limited to the observed data set, and thus cannot be generalised for larger populations.

With respect to the difference observed between the gender categorisations and the intercepts of the cue weights between the two interfaces, one could speculate that this could be the result of an interference effect: with more conflicting cues, participants have more difficulties answering. Such interference could be caused by the potential for participants to subconsciously attribute a gender to Sam based on their physical appearance despite our efforts to keep Sam gender neutral [66–70]. This may have occurred in this test based on the intercepts for F0. The value of the intercept, as determined by the logistic regression above, indicates the degree to which there is a bias in categorising a voice as a female gender with a 0 st difference to the reference voice. A lower intercept would indicate more of a bias toward male categorisations over female categorisations. Furthermore, this bias was significantly lower with Sam (0.90 ± 1.11) in comparison to the laptop (1.37 ± 0.95), showing a bias toward male categorisations with Sam.

It must be noted that although we attempted to implement this test with Sam as comparably similar to the laptop as possible, some elements from the laptop game-like implementation could not be replicated on Sam. In the laptop version, a cartoon image of a person was presented after the stimulus was presented to which participants could either agree or disagree on whether they perceived the voice gender to match the gender of the image. Although at the processing level the images had a specific gender associated with them, this binary choice was not as evident to the participants as with Sam, which presented only the right or left hands for male or female; images depicted on the laptop interface were not always stereotypically male or female. Based on the design of the test on the laptop, participants were asked to identify if the voice matched the image, not if the voice was male or female. Thus, whilst the test required participants to make a binary male or female categorisation on Sam, to some extent this choice was an abstraction in the laptop version. Additionally, using visual images of female and male faces on the computer was also not entirely bias-free as their evaluation as female or male would still rely on the participant's own concept of female and male, and how well the images would fit with these. Further, it could also be argued that there is no visual cue change with Sam; therefore, if some bias was present based on the physical appearance of Sam, it would have been consistent throughout the test. Hence, such consistent bias across stimuli should only lead to a shift of results with respect to the reference voice, but no further noise across trials. Different than in Experiment I, in Experiment II all these biases are inherent to both interfaces, potentially affecting the perceived voice gender when presented with face images or Sam, but what matters is that the test seems to be usable on each interface, producing meaningful data.

Comparing the observed data to that collected by [12], voice gender categorisation using Sam showed a statistically significant difference with respect to the intercepts [$t(19.858) = -2.887$, $p = 0.009$, Cohen's d = -1.060, $BF_{01} = 19.036$]; however, no statistically significant difference with respect to the F0 cue weights [$t(17.624) = -1.923$, $p = 0.07$, Cohen's d = -0.748, $BF_{01} = 2.600$], but a significant difference with respect to the VTL cue weights [$t(18.811) = -$ 2.751, $p$ = 0.013, Cohen's d = -1.055, $BF_{01}$ = 18.459]. Data obtained when using the laptop closely matches the voice gender categorisation of the adult group from data reported by [12], as depicted in Fig 6. Furthermore, there is no statistically significant difference between the intercepts of previous data and the laptop data [$t$(18.318) = -1.990, $p$ = 0.062, Cohen's d = -0.759, $BF_{01}$ = 2.763]; however, there was a statistically significant difference in the F0 cue weights between previous data and the laptop [$t$(17.359) = –2.311, $p$ = 0.033, Cohen's d = -0.906, $BF_{01}$ = 6.670]; and with respect to the VTL cue weights [$t$(20.517) = -2.375, $p$ = 0.027, Cohen's d = -0.859, $BF_{01}$ = 4.963].

Therefore, although variation was seen in the intercepts between the laptop and Sam, Sam seems to still produce data comparable with the laptop; however, both the laptop and Sam data showed varying degrees of differences when compared to the data reported by [12]. This can be interpreted as whether performing the voice gender categorisation test on the laptop or on Sam, as far as examining the influence of F0 and VTL on voice gender categorisation, there is no difference in consistency between the two interfaces.

We implemented the experiment such that the pairing of the gender and Sam's hand were randomised after each presented stimulus, and Sam would indicate which hand should be used for each gender. As expected the test took longer to complete on Sam in comparison to the laptop, however, this was due to other reasons than those in Experiment 1. Nevertheless, results for voice gender categorisation collected from Sam are still identical to those collected on the laptop. Furthermore, the increase in duration is also a consequence of the design of the implementation on Sam, as expected. After each stimulus is presented, Sam takes a few seconds to indicate which hand should be touched for a male or female categorisation, lengthening the duration of the test. This design choice was made to prevent the possibility of participants attributing a gender to one hand, and using that hand when voices were more difficult to categorise, potentially introducing an additional bias. In addition, test duration may have been increased due to a lack of visual cues when categorising a voice. The categorisation of a voice may be easier when presented with a face, as in the laptop version, thus making the decision process quicker. In comparison, participants had to pay more attention to the presented voice with Sam as this was the only cue presented, possibly causing hesitations in participants' categorisations, and increasing the duration of the test. The total duration of the test on Sam was also longer than that reported in previous studies, which corresponds to approximately three minutes [12] [$t$(35.706) = 13.18, $p <$ 0.001, Cohen's d = 3.160].

## General discussion

In both auditory research and clinical settings, due to the inherent repetitiveness and long duration of many auditory perception psychophysics tasks, using a humanoid NAO robot was suggested as a new alternative interface. If it can produce reliable data, an additional advantage could be to make the testing more enjoyable and help with engagement. Therefore, as a first step, the goal of the present experiments was to evaluate our NAO robot, Sam, as an auditory testing interface by comparing participants' results (in both test performance accuracy and duration) on voice perception and voice gender categorisation when using Sam to those when using the current laptop game-like interfaces and to previously reported literature.

### Comparison of test performance

The test performance was measured in JNDs (Experiment I) and Berkson units (Experiment II). Results have shown that, overall, Sam is comparable to the laptop interface and to previous studies that have used the same or similar test procedures, thus meaning the measured performances were similar within participants between both interfaces. Although some discrepancies were observed in the experiments, there was always some consistency either between the two interfaces or between Sam and previously reported data.

The small discrepancies observed in the results could have been due to a number of factors, such as differences in sound quality between the interfaces, as well as the use of English stimuli in comparison to the Dutch stimuli previously used [10, 12]. In addition, [12] used headphones, whereas in the two experiments of the present study, stimuli were presented by the internal speakers of Sam and the laptop. Without the use of specially designed speakers or headphones, it is possible that stimuli were presented sub-optimally and possibly with some degradation. It is not possible to connect external speakers or headphones to Sam, thus using these on the laptop would produce an unfair comparison. Despite these discrepancies, auditory test performance was shown to be similar between the two interfaces, as shown in Figs 2 and 4 for the voice cue sensitivity and voice gender categorisation, respectively.

Certain design choices made when implementing the tests onto Sam may have also affected the obtained test performance results. Most notably would be the visual cues provided by the laptop and not by Sam, such as the faces in Experiment II. Without these visual cues, it might be difficult or require some memorisation for the participants when logging responses on Sam, as the methods to do so are not as salient in comparison to the laptop. This is largely applicable to Experiment I, which required participants to remember how the order of presented stimuli related to the sensors on Sam.

The choices in design are a consequence of attempting to implement tests that had originally been designed to be run on a computer interface onto Sam. It is not yet clear how much of an impact such choices have on the test results and if they even pose limitations. With more variations of similar experiments in future studies, these details would be optimised.

For using a robot for voice perception tests, previous work by [71] explored how the pitch of a robot's voice in conjunction with its physical appearance is perceived by children and its potential influence on user acceptance. Subsequent to this, [72] used the NAO to investigate how synthesised voices to sound that of male or female impacted children's perception of the gender of the robot. The authors had hypothesised that children could attribute a male gender to the NAO regardless of the gender of the voice. Although the study does not explain the motivation for this specifically, based on other studies, this increased attribution of a male gender to the robot could be due to the robot's body shape more resembling a typical male body than female body [66, 67]. This generalised male attribution to the NAO by both children and adults is also commented on by [68]. In their study, school-age children interacted with the robot for an average of 10 minutes. The experiment was an interactive game wherein the robot, which either had a male or female voice, asked for a specific card from a selection lying on a table. The game itself was not relevant to the research question; however, following the game, children were interviewed on what gender they perceived the robot to have. Results showed that younger children (5–8 years) were less likely to assign a gender to the robot based on the perceived voice in comparison to older children (9–12 years), showing that younger children tended to attribute a gender to the robot independently of its voice. A potential gender categorisation bias could affect the performance of Sam as an auditory test interface, as Experiment II showed a bias towards male categorisations even when stimuli were presented at the reference female voice. While this bias needs to be further investigated to confirm if it exists, there is a possibility that despite the attempts at gender-neutrality when referring to Sam (such as always referring to it as "Sam" and avoiding the use of gendered pronouns when talking about the robot to participants), participants may have inadvertently attributed a male gender to Sam, potentially biasing their categorisations. In contrast, however, adults seem to assign a gender to a robot based on the voice, even when the robot is presented as gender neutral [69]. When the robot body is presented in a gender specific manner, adults seem to further enforce their stereotypical perception of the robot's gender [72, 73]. Further, adults even seem to have preferences for different voices for different task applications [74]. These results suggest that in identifying a voice presented via a humanoid robot, the perceived gender of the voice may be biassed based on the appearance of the robot and the context, and vice versa, potentially affecting the overall voice perception process. Since we have not explicitly studied this potential bias, which may also be part of the laptop version, we cannot yet conclude with certainty if one interface would have more bias than the other. Nevertheless, the potential gender attribution to Sam will be investigated in the follow-up human robot interaction study.

### Comparison of test durations

In so far as an auditory interface, we have shown Sam to be comparable to the laptop and previous similar studies with regard to the voice cue sensitivity and voice gender categorisation tests; however, it is also important to address the increased durations of the tests. In some cases, the longer test duration could be attributed to the design of the test implementation on Sam. One such factor that affected the increased duration was the stimulus processing time on Sam in Experiment I. Although a stimulus took on average two seconds to process on Sam, on the laptop it was around one second, on average resulting in a doubling of the time to complete the test on Sam. Having observed how much longer it takes to real-time process speech stimuli by the robot, for further applications and until there are substantial improvements to the robot processing power, using offline processed stimuli for a robot interface for auditory testing could be recommended (similar to Experiment II). In Experiment II, the increased duration of the test was due to Sam indicating which hand to use for the two gender categories (taking five seconds) after each of the 36 stimuli. This resulted in the presentation of stimuli at a slower rate than with the laptop, adding an extra three minutes to the total test duration.

In summary, despite the limitations due to the design of the implementation on Sam, it was observed that auditory perception testing results collected on Sam were still comparable to those from the laptop and previously reported data. While we will conduct a detailed analysis of the human robot interaction-related aspects, the observations of the present study support the potential for Sam to conduct longer auditory perception tests.

### A word on inferential statistics

The decision to use both classical (Frequentist) and Bayesian inference in this manuscript was primarily due to the design and expected outcome of the experiments. When using the classical p-value inferential approach, often the desired outcome is a small enough p-value to infer that any difference in the data is not due to chance but can also be extrapolated to a larger population. However, when the p-value is large, this is interpreted as insufficient certainty as to whether a difference exists. In comparison, because Bayesian inference is solely focussed on the observed data (and not based on a hypothetical data set as with classical inference) [56], it provides an alternative interpretation of the data; how much evidence (based on the observed data) can be attributed to the presence or absence of an effect.

However, it must be noted that for both statistical approaches in both experiments, due to the respective effect sizes and anecdotal evidence, results reported here are likely affected by the sample size. The benefit of using the Bayesian approach is that the current results can be used as the prior for follow-up tests using identical setups, thus providing stronger inferential power. In comparison, the use of the p-value (the Frequentist approach) would require either conducting the same experiment again with a new hypothesis and sample or redesigning of the experimental setup.

Examining the conducted experiments closer, the difference between the frequentist and Bayesian results regarding the gender categorisation between the interfaces shows one of the benefits of using the Bayesian approach. Based on frequentist methods, a statistically significant difference, without reporting an effect size (such as Cohen's d), would infer that the categorisation of voice gender is significantly dependent on the interface used. However, this effect size is included in the Bayesian result, and thus with a single statistic, it can be inferred that although there may be a difference in the results, there is only anecdotal evidence supporting it, and thus likely is limited to the observed data.

Comparing the results of both the classical and Bayesian approaches in the above experiments, consistency can be observed between statistical significance (classical) and evidence (Bayesian) based approaches. In the first experiment, the only discrepancy between the two approaches to the statistics was for any difference between the F0 JND thresholds obtained on either interface. The small effect size (0.084) and anecdotal evidence ($BF_{10}$ = 1.638) are both good indicators that based on the observed data, it cannot be concluded if there is indeed a difference between the two interfaces for this vocal cue, but also that there is no difference. Therefore, this would require further testing to appropriately draw a conclusion. Similarly, in Experiment II the only discrepancy between the classical and Bayesian statistics was the overall voice gender categorisation between the two interfaces. These results too had a small effect size (0.1) and low Bayes factor ($BF_{10}$ = 1.248), and thus again, no appropriate conclusion can be drawn as to the existence of a difference between the two interfaces.

## Concluding remarks

The auditory performance results of the present study show promise for conducting the PICKA psychophysics tasks, insofar as the voice cue sensitivity and voice gender categorisation tests are concerned on Sam. Furthermore, our results contribute to the growing potential of using humanoid robots for both learning and testing applications [34–36, 75, 76], also for specific target groups and rehabilitation applications (e.g., [43, 77, 78]), and our general understanding of speech communication and voice perception in HRI, an increasingly relevant topic for social robots [1, 13, 79, 80].

To not only further reduce the duration of the tests on Sam, but also to negate possibility confounding effects due to inconsistent positive and negative feedback between the two interfaces [81], the visual feedback during the voice cue sensitivity test could be simplified such that only positive feedback is presented (nodding only when the given response is correct); otherwise, continuing with the next stimulus, shortening the test duration. Although this feedback only takes one or two seconds, as mentioned with the hand identification during the gender categorisation test, the summation of these delays after each stimulus can result in a longer than intended test.

An additional, albeit more complex, technical modification that could be made to the implementation of both tests could be the incorporation of automatic speech recognition (ASR). Rather than logging responses by touching Sam's sensors, verbal responses could be given. Although the NAO does have speech recognition capabilities, it is not expected that it would be sufficient for such testing purposes, as the speech recognition module performs poorly and often requires higher than natural speech volumes in order to recognise any speech. However, should this problem be overcome through the use of third-party software, such as the speech recognition system Kaldi [82], this could be a viable alternative to the tactile response logging method.

## Acknowledgments

We thank Josephine Marriage and Debi Vickers for English stimuli, Paolo Toffanin, Iris van Bommel, Evelien Birza, Jacqueline Libert, and Jop Luberti for their contribution to the development of the PICKA test battery. Parts of the study were conducted as master's thesis projects by Thirsa Huisman (supervised by Leanne Nagels), and Luke Meyer.

## Author Contributions

**Conceptualization:** Luke Meyer, Etienne Gaudrain, Deniz Başkent.

**Data curation:** Luke Meyer.

**Formal analysis:** Luke Meyer, Etienne Gaudrain.

**Funding acquisition:** Luke Meyer, Deniz Başkent.

**Investigation:** Luke Meyer, Laura Rachman.

**Methodology:** Luke Meyer, Laura Rachman, Etienne Gaudrain, Deniz Başkent.

**Project administration:** Luke Meyer, Deniz Başkent.

**Resources:** Luke Meyer, Laura Rachman, Deniz Başkent.

**Software:** Luke Meyer.

**Supervision:** Laura Rachman, Gloria Araiza-Illan, Deniz Başkent.

**Visualization:** Luke Meyer, Etienne Gaudrain, Deniz Başkent.

**Writing – original draft:** Luke Meyer.

**Writing – review & editing:** Luke Meyer, Laura Rachman, Gloria Araiza-Illan, Etienne Gaudrain, Deniz Başkent.